\documentclass[10pt]{iopart}
\usepackage{geometry}                
\usepackage[parfill]{parskip}    
\usepackage{graphicx}
\usepackage{amssymb}
\usepackage{epsfig}
\usepackage{epsf}
\usepackage{epstopdf}
\begin{document}
\title{Self Similar Renormalization Group Applied to Diffusion in non-Gaussian Potentials }
\author{David S. Dean and Cl\'ement Touya}

\address{Laboratoire de Physique Th\'eorique, Universit\'e Paul Sabatier, Toulouse, France}

\begin{abstract}
We study the problem of the computation of the effective diffusion constant of a Brownian particle
diffusing in a random potential  which is given by a function $V(\phi)$  of a Gaussian field $\phi$.
A self similar renormalization group analysis is applied to a mathematically related problem of
the effective permeability of a random porous medium from which the diffusion constant of the 
random potential problem can be extracted. This renormalization group approach reproduces
practically all  known exact results in one and two dimensions. The results are confronted with numerical simulations and we find that their accuracy is good up to points well beyond the expected
perturbative regime. The results obtained are also tentatively applied to interacting particle systems
without disorder and we obtain expressions for the self-diffusion constant in terms of the excess thermodynamic entropy. This result is of a form  that has commonly been used  to fit the self diffusion 
constant in molecular dynamics simulations. 
\end{abstract}
\maketitle

\section{Introduction}
In this paper we will consider the late time diffusion constant associated with a Brownian or Langevin
particle, in D dimensions, advected by a velocity field which is given by the gradient of a random potential. The 
explicit Langevin equation studied is:	
\begin{equation}
{d\mathbf{X}\over  dt} = \nabla V\left(\phi(\mathbf{X})\right) + {\mathbf{\eta}}(t) \label{sde}
\end{equation}
where the local potential $V$ is itself a function of a Gaussian random field $\phi$.  In general (up to an
overall rescaling of time)  the 
potential $V$ can be written and 
\begin{equation}
V(\phi) = -\beta \nu(\phi)
\end{equation}
where $\beta$ is the inverse temperature and $\nu$ is the physical potential acting on the tracer
particle. In this formulation $\eta(t)$ is a Gaussian white noise of zero mean with correlation function
\begin{equation}
\langle \eta_i(t)\eta_j(t')\rangle = 2\delta_{ij}\delta(t-t').
\end{equation}
Therefore when $V$ is not a linear function of $\phi$ the advecting potential is non-Gaussian. The 
case of diffusion in purely Gaussian potentials has been extensively studied in the literature \cite{ddh,dech,rev} but the non-Gaussian case has received much less attention. 
The Gaussian case has been studied
within a variety of approximation schemes and among these schemes the most successful has been
the self-similar renormalization group method which reproduces exact results in one and two dimensions and which in addition is in excellent agreement with numerical simulations in three
dimensions \cite{ddh,dech,rev}. The case where $V$ is the square of the gradient of a Gaussian potential arises naturally in the case of dipoles diffusing in a random electric field (in the limit where the dipole moment equilibriates very
quickly in its local field compared to the time-scales over which the diffusion of its centre of mass 
occurs) \cite{dhs}. The authors of this paper have examined the case 
\begin{equation}
\nu = \alpha {\phi^2\over 2}
\end{equation}
in one dimension \cite{tode} , where the diffusion constant can be calculated exactly. In this case it can be shown that there is a critical temperature at which the diffusion constant vanishes. Below this temperature the diffusion is anomalous (more precisely sub-diffusive) and the exponent associated with anomalous diffusion can be computed. 
The exact results of \cite{tode} show that the transport properties (the exponent associated with the anomalous diffusion) agree with those obtained by a straightforward mapping onto a trap model whose
trapping time statistics can be deduced from the statistics of the field $\phi$ and the Arrhenius law.
There have also been some studies of diffusion in (non-Gaussian) potentials generated by the 
potentials due to a distribution of randomly distributed particles which interact with the potential 
via a fixed deterministic potential \cite{ddh3,ddhl}.

A naive application of the self-similar renormalization group to this model at the one-loop level
is not sensitive to the non-Gaussian statistics of the random potential and fails to predict the 
dynamical phase transition associated with the passage from normal to sub-diffusive transport. In this
paper we reformulate the self-similar renormalization group approach in such away that the results for
the Gaussian case are unchanged but we reproduce all known exact results in one and two dimensions. In addition we show that the approach works well in other cases by comparison with 
numerical simulations, it predicts the dynamical transition, when there is one, and works reasonably
well outside the perturbative regime. The basis of our analysis relies on the mapping of diffusion in the
random potential to diffusion in a medium of random diffusivity, this is mainly done as it simplifies the
resulting renormalization group flow. Interestingly, as a biproduct our analysis recovers an approximate results often used in the computation of the effective permeability of random porous media 
\cite{rev,mat,king,spo,hir,eber}

The underlying Gaussian potential we shall study will be assumed to have a short range correlation 
of the from
\begin{equation}
\langle \phi(\mathbf{x})\phi(\mathbf{y}) \rangle = \Delta(\mathbf{x}-\mathbf{y})
\end{equation}

In a translationally invariant and isotropic system the long-time behavior of the mean-squared displacement of the process $\mathbf{X}$ described by equation (\ref{sde}) is  
\begin{equation}
\langle \mathbf{X}^2\rangle\sim 2D\kappa_e t 
\end{equation}
where $\kappa_e$ is thus the long-time effective diffusion constant of the problem.
The Fokker-Planck equation describing the evolution of the probability density function 
(pdf) for $ \mathbf{X}$
is:
\begin{equation}
{\partial P\over \partial t} = \nabla\cdot\left(\nabla P -P\nabla V(\phi) \right) \label{fpg}
\end{equation}
The problem of diffusion in a medium of random local diffusivity $\kappa= \exp(V(\phi))$ is
described by the Fokker-Planck equation 
\begin{equation}
{\partial P\over \partial t} = \nabla\cdot \exp\left( V(\phi)\right)\nabla P  \label{fpp}.
\end{equation}
The corresponding stochastic differential equation is
\begin{equation}
{d\mathbf{X}\over  dt} = \exp\left(V\left(\phi(\mathbf{X})\right)\right) \nabla V\left(\phi(\mathbf{X})\right) + \sqrt{\exp\left(V\left(\phi(\mathbf{X})\right)\right)}{\mathbf{\eta}}(t) \label{sdek}.
\end{equation}
Again under the assumptions of time-translational invariance and isotropy the late-time behaviour
of a Brownian tracer particle described by the above Fokker-Planck equation is 
\begin{equation}
\langle \mathbf{X}^2\rangle\sim 2D\kappa_e^{(p)} t 
\end{equation}
where $\kappa_p$ is the associated long-time diffusion constant. The effective diffusion constant
$\kappa_e^{(p)}$ is also the effective permeability of a random porous medium where fluid flow is
described by Darcy's law and if $\kappa$ is interpreted as a local dielectric constant then $\kappa^{(p)}_e $ is the effective dielectric constant of the medium (see the review \cite{rev}). The effective late-time  diffusion constants of the two Fokker-Planck equations ({\ref{fpg}) (gradient flow) and ({\ref{fpp})
(fluctuating diffusivity) 
are in fact related via
\begin{equation}
\kappa = {\kappa_p\over \langle \exp\left(V(\phi)\right)\rangle },\label{kpkgr}
\end{equation}
a result which can be shown in a number of different ways \cite{rev,ddhl,ddh2}. The effective diffusivity 
can be computed from a statics problem. If one considers the Green's function for the random diffusivity
problem 
\begin{equation}
\nabla\cdot \exp\left(V(\phi)\right)\nabla G   = -\delta(\mathbf{x}),\label{gfp}
\end{equation}
the effective diffusivity can be read off from the long-distance behavior of the Green's function or
equivalently the short wave-length behaviour of its Fourier transform: 
\begin{equation}
\langle {\tilde G}(\mathbf{k})\rangle  \sim {1\over \kappa_e^{(p)} \mathbf{k}^2}
\end{equation}
which means that on suitably large length scales  Green's function reads
\begin{equation}
 G ^{-1} \sim -\kappa_e^{(p)}\nabla^2.
\end{equation}
Note we have dropped the disorder average as we shall assume that $\kappa_{e}^{(p)} $is self-averaging.

\section{Renormalization Group Approach}

The basic idea of the self-similar renormalization group \cite{ddh,dech} is to average out the short distance 
components of the random field $\phi$ down to some wave-length $\Lambda$ and then to write
an effective diffusion equation, of the same structural form, describing the transport at length scales
greater than $1/\Lambda$.

The Gaussian field is decomposed by defining
\begin{equation}
\phi_\Lambda(\mathbf{x}) = \int_{|\mathbf{k}|<\Lambda} {d\mathbf{k}\over (2\pi)^D}
\exp(i\mathbf{k}\cdot \mathbf{x})\tilde\phi(\mathbf{k})
\end{equation}
where $\Lambda$ is an upper ultraviolet scale which is initially infinite. We define
a slice of the field $\phi$ at inverse length scale $\Lambda$
\begin{equation}
\delta \phi_\Lambda(\mathbf{x}) = \int_{\Lambda-\delta\Lambda|\mathbf{k}|<\Lambda} {d\mathbf{k}\over (2\pi)^D}\exp(i\mathbf{k}\cdot \mathbf{x})\tilde\phi(\mathbf{k})
\end{equation}
The self-similar renormalization group process proceeds by integrating out the slice of the field
$\delta\phi$ whilst assuming that the remaining part of the field $\phi_{\Lambda-\delta\Lambda}$ 
can be treated as a constant at the length scale $\Lambda$. The correlation function of the slice
of the field $\delta\phi_\Lambda$ is given by
\begin{equation}
\langle \delta \phi_\Lambda(\mathbf{x})\delta \phi_\Lambda(\mathbf{y})\rangle 
= \Delta_{\delta\Lambda}(\mathbf{x}-\mathbf{y})
\end{equation}
and its Fourier transform is given by
\begin{equation}
\tilde \Delta_{\delta\Lambda}(\mathbf{k}) = \tilde \Delta(\mathbf{k}) I(|\mathbf{k}|, [\Lambda-\delta\Lambda
,\Lambda])
\end{equation}
where $I$ is the indicator function
\begin{eqnarray}
I(x,A) &=& 1\ \ {\rm if}\ x\in A \nonumber \\
&=& 0 \ \ {\rm if}\ x\notin A.
\end{eqnarray}
The field $\delta\phi_\Lambda$ is thus formally of order $\sqrt{\delta\Lambda}$. This means that
to order $\delta\Lambda$ one may write at any given point
\begin{eqnarray}
\delta \phi_\Lambda^2 &=&\langle \delta \phi_\Lambda^2\rangle \nonumber \\
                                          &=& \delta \mu = {S_D\Lambda^{D-1} \tilde\Delta(\Lambda)\over (2\pi)^D}\delta\Lambda 
\end{eqnarray}
If we apply the self-similar renormalization group hypothesis to the Green's function in equation
(\ref{gfp}) we expect that after integrating out the random field down to wave number $\Lambda$ that
on this inverse length scale the running Green's function obeys a similar renormalized equation of
the form 

\begin{equation}
\nabla \cdot \exp(V_\Lambda (\phi_\Lambda ))\nabla G_\Lambda = -\delta(\mathbf{x}).
\end{equation}
Here $G_\Lambda$ denotes the Green's function averaged over modes of $\phi$ of modulus
superior to $\Lambda$ which we denote as:
\begin{equation}
G_{\Lambda} = \langle G\rangle_\Lambda
\end{equation}
Now in this equation we write the field $\phi_\Lambda = \phi_{\Lambda-\delta\Lambda} + \delta\phi_{\Lambda}$. In the running equation for $G_\Lambda$ we treat $\phi_{\Lambda-\delta\Lambda}$ as 
approximately constant to obtain
\begin{eqnarray}
&&\exp\left(V_\Lambda(\phi_{\Lambda-\delta\Lambda}) + {\delta\mu\over 2}\left(V_\Lambda^{''}(\phi_{\Lambda-\delta\Lambda})  + (V^{'}_\Lambda(\phi_{\Lambda-\delta\Lambda}))^2\right)\right) \nonumber \\
&&\nabla\cdot (1+ V^{'}_\Lambda(\phi_{\Lambda-\delta\Lambda})\delta\phi_\Lambda)\nabla G_\Lambda
= -\delta(\mathbf{x}).
\end{eqnarray}
Under these hypotheses we obtain:
\begin{equation}
G_\Lambda = {g_{\Lambda}\over \exp\left(V_\Lambda(\phi_{\Lambda-\delta\Lambda}) + {\delta\mu\over 2}\left(V_\Lambda^{''}(\phi_{\Lambda-\delta\Lambda})  + (V^{'}_\Lambda(\phi_{\Lambda-\delta\Lambda}))^2\right)\right)},
\end{equation}
where the Green's function $g_\Lambda$ is defined via
\begin{equation}
\nabla\cdot (1+ V^{'}_\Lambda(\phi_{\Lambda-\delta\Lambda})\delta\phi_\Lambda)\nabla g_\Lambda.
= -\delta(\mathbf{x})
\end{equation}
Now averaging over the current momentum slice we find that
\begin{equation}
G_{\Lambda-\delta\Lambda} = {\langle g_{\Lambda}\rangle_{\delta\Lambda}\over \exp\left(V_\Lambda(\phi_{\Lambda-\delta\Lambda}) + {\delta\mu\over 2}\left(V_\Lambda^{''}(\phi_{\Lambda-\delta\Lambda})  + (V^{'}_\Lambda(\phi_{\Lambda-\delta\Lambda}))^2\right)\right)}.
\end{equation}
Note that the average  over the slice of the field $\delta\phi_\Lambda$ can be carried out in the 
computation of $\langle g_\Lambda\rangle$. Now at length scales $\Lambda$ the Green's function $g_\Lambda$ 
should  behave as
\begin{equation}
\langle \tilde g_{\Lambda}(\mathbf{k})\rangle \sim {1\over \kappa^{*}(\phi_{\Lambda-\delta\Lambda}) \mathbf{k}^2}.
\end{equation}
The equation determining $g_\Lambda$ is of the form
 \begin{equation}
\nabla\cdot (1+ \psi)\nabla g \label{gfgl}
= -\delta(\mathbf{x})
\end{equation}
where $\psi$ is a Gaussian field with correlation function
\begin{equation}
\langle \psi(\mathbf{x})\psi (\mathbf{y})\rangle = D(\mathbf{x}-\mathbf{y}).
\end{equation}
Taking the Fourier transform of equation (\ref{gfgl}) yields
\begin{equation}
\tilde g(\mathbf{k}) = {1\over \mathbf{k}^2} - \int {d\mathbf{q} \over (2\pi)^D} \mathbf{k}\cdot(\mathbf{k}-
\mathbf{q})\tilde \psi(\mathbf{q})\tilde g(\mathbf{k}-\mathbf{q}).
\end{equation}
This equation can be iterated diagrammatically 
and can then be averaged to yield a set of Feynman diagrams which can be summed in terms of
one-particle irreducible diagrams to write:
\begin{equation}
 \tilde g(\mathbf{k})= {1\over  \mathbf{k}^2 - \Sigma(\mathbf{k})}
\end{equation}
At order $\delta\Lambda$ (which is simply one loop as the momentum in each loop is
$\delta \Lambda$) we find 
\begin{equation}
\Sigma(\mathbf{k}) = \int {d\mathbf{q} \over (2\pi)^D} \tilde D(\mathbf{q}){(\mathbf{k}\cdot(\mathbf{k}-
\mathbf{q}))^2\over (\mathbf{k}-
\mathbf{q})^2}\approx {\mathbf{k}^2\over D}\int {d\mathbf{q} \over (2\pi)^D} \tilde D(\mathbf{q})
\end{equation}
for small $|\mathbf{k}|$. Note that in principle we have introduced higher order derivatives and interactions and so the approach is clearly not exact. However we will see that this approach appears
to be capturing the essential physics of the problem. The correlation function here is given by
\begin{equation}
\tilde D(\mathbf{q}) = {V'}_\Lambda^{2}(\phi_{\Lambda-\delta\Lambda})\tilde \Delta(\Lambda) I(\mathbf{q}, [\Lambda-\delta\Lambda, \Lambda])
\end{equation}
and thus we find that 
\begin{equation}
\kappa^{*}(\phi_{\Lambda-\delta\Lambda}) = 1 - {{V'}_{\Lambda}^{2}(\phi_{\Lambda-\delta\Lambda})\over D}\delta\mu.
\end{equation}
This yields 
\begin{equation}
\tilde G_{\Lambda-\delta\Lambda} \sim  {1\over \mathbf{k}^2 \exp\left(V_\Lambda(\phi_{\Lambda-\delta\Lambda}) + {\delta\mu\over 2}\left(V_\Lambda^{''}(\phi_{\Lambda-\delta\Lambda})  + (1-{2\over D}){V'}_\Lambda^2(\phi_{\Lambda-\delta\Lambda})\right)\right)}
\end{equation}
we now associate the prefactor of the term $\mathbf{k}^2$ in the denominator above as the effective
diffusion constant in a region of size $1/\Lambda$ which we denote as $\exp\left(V_{\Lambda}(\phi_{\Lambda-\delta\Lambda}\right)$. We now compute the flow of the function $V_\Lambda$ to obtain
\begin{equation}
{\partial V_\Lambda\over \partial \Lambda} = {1\over 2}{d\mu\over d\Lambda}\left(V_\Lambda^{''} + \left(1-{2\over D}\right){V'}_\Lambda^2\right)\label{rgflow}
\end{equation}
The boundary conditions on $V_\Lambda$ is $V_{\infty} = V$ and the 
effective diffusion constant is given as 
\begin{equation}
\kappa_e^{(p)} = \exp(V_0(0))
\end{equation}
{\em i.e.} after all the random modes have been integrated out. The renormalization group flow
equation is non-linear but in the case where $V = a\phi^2 + b\phi + c$ the flow does not introduce
new interactions and the full solution can be computed. However one may formally compute 
the effective diffusion constant via the following observation. If one wants to compute the average
\begin{equation}
A_0 = \langle \exp\left(\alpha U(\phi)\right)\rangle
\end{equation}
one may also use a (albeit very simple) renormalization group procedure writing
\begin{eqnarray}
A_\Lambda &=& \langle  \exp\left(\alpha U(\phi)\right)\rangle_\Lambda\nonumber \\
 &=&\exp\left(\alpha U_{\Lambda}(\phi_{\Lambda})\right).
\end{eqnarray}
The flow equation for $U_\Lambda$ is easy to compute and is given by
\begin{equation}
{\partial U_\Lambda\over \partial \Lambda} = {1\over 2}{d\mu\over d\Lambda}\left(U_\Lambda^{''} + \alpha{U'}_\Lambda^2\right)
\end{equation}
Thus if we make the following identification 
\begin{equation}
V_\infty =U_\infty\ \ ; \ \ \alpha = 1-{2\over D}
\end{equation}
we find that $V_\Lambda= U_\Lambda$ for all $\Lambda$ and consequently
\begin{eqnarray}
\kappa_{e}^{(p)} = \exp\left(V_0(0)\right) &=& \exp\left( \left(1-{2\over D}\right)V_0(0)\right)^{1\over \left(1-{2\over D}\right)}
\nonumber \\
&=& \left\langle \exp\left(\left(1-{2\over D}\right)V(\phi)\right)\right\rangle^{1\over \left(1-{2\over D}\right)}. \label{llm1}
\end{eqnarray}
Therefore if one has a local diffusivity or permeability $\kappa(\mathbf{x}) = \kappa(\phi(\mathbf{x}))$
i.e. that is an arbitrary (positive) function of a Gaussian field, then we find the effective 
permeability is given by: 
\begin{equation} 
\kappa_e^{(p)} =\left\langle\kappa(\mathbf{x})^{1-{2\over D}}\right\rangle^{1\over \left(1-{2\over D}\right)}\label{llm}
\end{equation}
This result is a widely used approximation in the field of effective permeabilities and this form is sometimes referred to as the Landau-Lifshitz-Matheron conjecture \cite{mat,hir}
(although is is usually stated
for Gaussian fields in terms of the local field variance). This formula is exact in one dimension
and is also exact in two dimensions if the local permeability is (up to a constant multiplicative
factor) statistically identical to its inverse \cite{rev,ddh3,dyk1,dyk2}. If one repeats the argument 
above for a system where
\begin{equation}
\kappa = \exp(V(\phi) -V(\phi'))
\end{equation}
where $\phi$ has the same statistics as $\phi'$ then we find that $\kappa_e^{(p)}=1$ which 
agrees with the exact result in two dimensions. The result is not exact for the Gaussian 
case $V(\phi)=\phi$ in three dimensions but the deviation from the real result in fact only shows
up at three loop order \cite{dew}. In the hydrology community the question whether the  Landau-Lifshitz-Matheron conjecture was exact in three dimensions animated debate for sometime. 

Now we return to the problem of diffusion advected by the gradient of 
random potential $V(\phi)$, putting together the results of approximate equation (\ref{llm1})
and the exact relation (\ref{kpkgr}) we obtain
\begin{equation}
\kappa_e = {\left\langle \exp\left((1-{2\over D})V(\phi)\right)\right\rangle^{1\over \left(1-{2\over D}\right)}
\over \left\langle \exp\left(V(\phi)\right)\right\rangle}.\label{eqmr}
\end{equation}
This is the main result of our paper and in what follows we shall analyse the behaviour
for various choices of the potential $V(\phi)$ and confront the predictions with results of numerical simulations.

\section{Discussion and some special cases} 

In the case of a purely Gaussian potential $V(\phi) = -\beta \phi$ this gives
\begin{equation}
\kappa_e = \exp\left(-{\beta^2\over D}\right), \label{kgas}
\end{equation}
where we have set the variance of the Gaussian field $\Delta(0) =1$.
This results is in agreement with the renormalization group approaches in references \cite{ddh,dech}. It is 
known to be exact in one and  two dimensions and it is correct at two-loop order in perturbation theory. However  it has been shown to break down at  three loop order in three dimensions \cite{dew}
and so the result equation (\ref{eqmr}) is certainly not exact. However numerical simulations 
in three dimensions have shown that the prediction (\ref{kgas}) is remarkably accurate well 
beyond the perturbative region (where basic perturbation theory should work well). 

A case recently studied by the authors is that where we take
\begin{equation}
V(\phi) = -\beta{\phi^2\over 2},
\end{equation} 
{\em i.e.} the physical potential $\nu(\phi) = \phi^2/2$. 
Unlike the Gaussian case the diffusive behavior will depend on the sign of the inverse temperature
$\beta$. When $\beta$ is positive the Langevin particle is attracted to regions where $\phi = 0$. In
a $D$ dimensional space the regions where $\phi=0$ form  $D-1$ dimensional subspaces, at
low temperatures one thus expects the particle to be confined to these regions. However there
is no clear mechanism for confining the particle and thus we expect the transport to be diffusive at
all finite temperatures. When $\beta$ is negative the particle is attracted to points where the field $\phi$
is maximal or minimal where $\nabla \phi=0$. 

In the generic case where  attractive regions are zero dimensional 
they correspond to localised traps. In this case the average time to escape from a trap is given by the
Arrhenius law 
\begin{equation}
\tau \sim \tau_0 \exp(\beta \Delta E)
\end{equation}
where $\Delta E$ is the energy barrier associated with the trap and $\tau_0$ a microscopic time scale.  Now we will assume that 
effective energy barriers scale like  the potential $V$ itself and hence write    
$\Delta E =  \overline{\nu}-\nu({\phi})$ where $\overline \nu$ represents an arbitrary  energy level
at which one is deemed to be not trapped. This gives the mean residence time of a trap averaged over
traps to be 
\begin{equation}
\langle \tau \rangle \sim \tau_0 \exp(\beta \overline \nu)\left\langle\exp(-\beta\nu)\right\rangle 
\end{equation}
It is now easy to see that the average number of jumps from trap to trap is given by
$n\sim 1/\langle \tau\rangle$ and thus we find that 
\begin{equation}
\kappa_e \sim {1\over \tau} \sim {1\over \left\langle \exp(-\beta\nu)\right\rangle }\label{barrier}
\end{equation}
In terms of the physical potential $\nu$ our main result equation (\ref{eqmr}) reads
\begin{equation}
\kappa_e = {\left\langle \exp\left(-\beta\left(1-{2\over D}\right)\nu(\phi)\right)\right\rangle^{1\over \left(1-{2\over D}\right)}
\over \left\langle \exp\left(-\beta\nu(\phi)\right)\right\rangle}.\label{eqmr2}
\end{equation}
Comparing equations (\ref{barrier}) and (\ref{eqmr2}) we see the appearance on the same average 
$\left\langle \exp\left(-\beta\nu(\phi)\right)\right\rangle$ in the denominator; it is the divergence of this
term which is thus responsible for the vanishing of the diffusion constant. Let us note that,
even though this term may diverge at a certain value  $\beta=\beta_c$, the term in the numerator
which is the effective diffusion constant for the effective permeability {\em i.e.}
\begin{equation}
\kappa_e^{(p)} =\left\langle \exp\left(-\beta\left(1-{2\over D}\right)\nu(\phi)\right)\right\rangle^{1\over \left(1-{2\over D}\right)}\label{eqkep}
\end{equation}
remains finite beyond this value of $\beta$ and thus the random diffusivity problem can have a finite
diffusion constant while the gradient flow problem exhibits a vanishing diffusion constant. The 
localisation of  a dynamical transition, characterised  by a vanishing diffusion constant, via numerical simulations is notoriously difficult. First the low but finite value of the diffusion constant as one 
approaches the transition means that one must carry out simulations over long time scales
to diffuse sufficiently to place one in the steady state (time translationally invariant regime) and 
in order to reach the long time regime of the diffusion process. Furthermore, it was shown by
the authors \cite{tode}, that there are finite size effects if one uses a finite number of modes
to simulate the random field \`a la Kraichnan \cite{kra},  these effects smooth out the dynamical transition
in a similar way to which finite size effects affect simulations of critical phenomena. It is for this 
reason that it is sometimes better to simulate the random diffusivity problem, corresponding
to the gradient flow problem, and then deduce the effective diffusion constant for the gradient flow
problem via the exact relation equation (\ref{kpkgr}).  We will show a numerical example of 
the effectiveness of this approach later on. 

A specific example where the effective diffusion constant can be explicitly evaluated is
\begin{equation}
\nu(\phi) = {1\over 2}(\phi-a)^2,
\end{equation}
here we find
\begin{equation}
\kappa_e = {\left(1+\beta\right)^{1\over 2}\over \left(1 +\beta\left(1-{2\over D}\right)\right)^{{1\over  2\left(1-{2\over D}\right)}}}
\exp\left(-{\beta^2a^2\over D}{1\over \left(1+\beta(1-{2\over D})\right)(1+\beta)}\right).\label{eqke}
\end{equation}
In the case $D=1$ we find
\begin{equation}
\kappa_e = \left(1-\beta^2\right)^{1\over 2}
\exp\left(-{\beta^2a^2\over (1-\beta^2)}\right).
\end{equation}
which is an exact result. For $D=2$ we find
\begin{equation}
\kappa_e = (1+\beta)^{1\over 2}\exp\left(-{\beta\over 2}\right)
\exp\left(-{\beta^2a^2\over 2}{1\over (1+\beta)}\right).
\end{equation}
We note that for $D=1$ there is a transition for both positive and negative $\beta$. However
in higher dimensions there is only a transition predicted for $\beta$ negative ($\beta=-1$), that is to say when local maxima and minima of the field $\phi$ behave as traps. In the case where $a=0$
the diffusion constant vanishes in a power law fashion reminiscent to that predicted by 
mode coupling type theories \cite{mct}. In the case where $a\neq 0$ the dominant behavior in the vanishing
of the diffusion constant (or the divergence in the characteristic time scale) has the form
\begin{equation}
\kappa_e \sim \exp\left(-{C\over |T-T_c |}\right)
\end{equation}
which has the Vogel-Fulcher-Tammann form often evoked in the analysis of the experimental 
glass transition. Let us note that in one dimension 
the result for the diffusion constant is a function of $\beta^2$ and there is a dynamical transition at $\beta=1$ and $-1$. The transition at $\beta=-1$ is of course already guaranteed due to our formulation
of the problem via the diffusivity representation; the transition at  $\beta=1$ is however predicted
directly by the renormalization group analysis. 

The behaviour of $\kappa_e$ at positive $\beta$ in two and higher dimensions is very interesting.
Recall here the particle will be localised on the $D-1$ dimensional surfaces $\phi=a$ at low
temperatures. For large $\beta$ the renormalization group prediction is
\begin{equation}
\kappa_e \sim T^{{1\over D-2}}\exp\left(-{a^2\over D-2}\right)\left(1-{2\over D}\right)^{-{1\over 2\left(1-{2\over D}\right)}}
\end{equation}  
for $D>2$ and
\begin{equation}
\kappa_e \sim {1\over T^{1\over 2}}\exp\left(-{1\over 2T}\left(1+a^2\right)\right)
\end{equation}
for $D=2$.
We thus see that the dimension plays  a crucial role in the low temperature behaviour of the
diffusion constant, For $D=2$ the dominant effect of temperature on the diffusion constant is
of the Arrhenius form, implying that the crossing of energy barriers is the major mechanism 
contributing to diffusion. However for more than two dimensions there is a simple power law
behavior of the diffusion constant at low temperature, implying that the effect of energy barriers
is somehow marginal. This implies  that most transport can be achieved without crossing
energy barriers which are of order 1 and that the particle manages to diffuse while staying close
to the surface $\phi=a$.  It is worth remarking here that the way in which the Volger-Fulcher-Tammann
law arises here is identical to the mechanism arising in the glass model of Vilgis \cite{vil}. Here the 
local energy barriers are taken to be of the form $Ez$, where $E$ is the typical energy barrier due to 
a neighbouring  atom or molecule in a network type glass, and $z$ is the local coordination number.
The  energy barrier is thus $Ez$, and if one assumes that $z$ has a Gaussian distribution about an average value $z_0$ then one easily finds, via the Arrhenius law, the VFT form for the relaxation time.   

\section{Numerical simulations}
In this section we test the predictions of the renormalization group analysis against numerical 
simulations of the Langevin equations (\ref{sde}) and (\ref{sdek}). In what follows we will generate the random
fields using the method due to Kraichnan \cite{rev,kra}. In the simulations carried out here 
we found that away from any dynamical transition that the results were not significantly changed 
in going between 64 and 128 modes. The results reported  here will be in the majority of cases
for 128 modes. The stochastic differential equations for both the gradient flow and random 
diffusivity problem were integrated using second order Runga-Kutta integration schemes 
developed in \cite{dhh,hon} and reviewed in \cite{rev}. In all simulations the effective 
diffusion constant for a 
given realisation of the disorder was obtained by fitting the mean squared displacement averaged
over $10000$ particles at late times.  The time of the simulation was chosen so that particles had
typically diffused ten or so correlation lengths of the field. The diffusion constant is 
determined by a  fit of the average mean square displacement over the last half of the 
time of the simulation (to ensure that the mean squared displacement is well within the linear regime).
In three and higher dimensions the late time average mean square displacement is fitted with a 
simple linear fit  of the form $At + B$ and in two dimensions a logarithmic correction is
used {\em i.e.} $At + B \ln(t)$. 
Finally, the average over the disorder induced by the random field was made over $500$ realisations of the field. In all simulations the Gaussian field was 
taken to have correlation function
\begin{equation}
\Delta({\mathbf x}) = \exp\left(-{1\over 2}\mathbf{x}^2\right)
\end{equation} 
and we   concentrates on quadratic forms for the potential of the form  
$\nu = (\phi-a)^2/2$. Firstly we carried out numerical simulations of the diffusivity problem 
described by the Langevin equation (\ref{sdek}). Shown in figure ({\ref{kep}) is the numerically measured
value for the diffusion constant in two, three and four dimensions compared with that given by the 
renormalization group prediction for the case $a=0$. 
We see that the agreement is excellent up to very large values
of $|\beta|$ showing that the RG approach works well outside the expected perturbabitive regime.
The RG prediction appears to improve as the dimensions of the space is increased. In figure 
(\ref{kepa0.5}) we show the corresponding curves for the case $a=0.5$, again we see excellent agreement.

\begin{figure}
\begin{center}
\epsfig{file=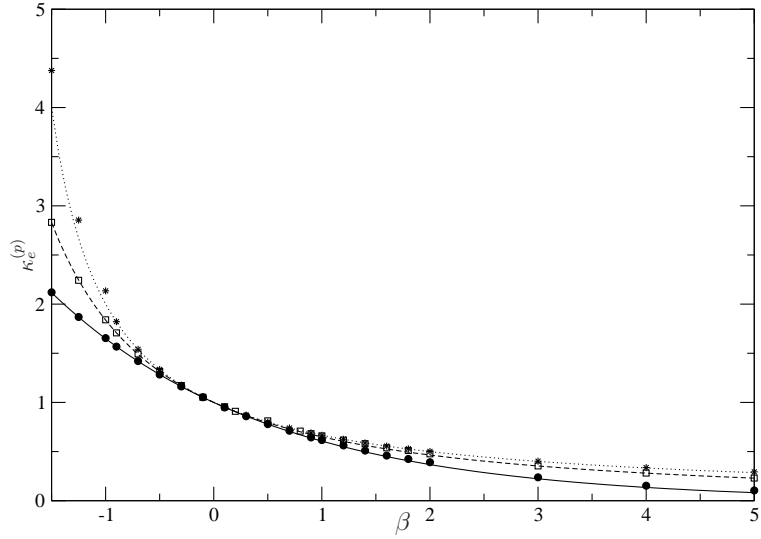}
\end{center} 
\caption{Analytical calculation of equation (\ref{eqkep}) for $\kappa_e^{(p)}$ for $a=0$ in 2(solid line), 
3 (dashed line) and 4 (dotted line) dimensions, confronted with direct simulation of the random diffusivity problem. 
The simulation results are shown as circles ( $2D$), squares  ($3D$) and stars for (4D). We have excellent agreement except at negative $\beta$ except for $\beta<-1$ in $4D$.
This is to be expected as locally $\kappa({\mathbf{x})}$ can become very big which is obviously a problem for simulation.}
\label{kep}
\end{figure}

\begin{figure}
\begin{center}
\epsfxsize=0.6\hsize\epsfbox{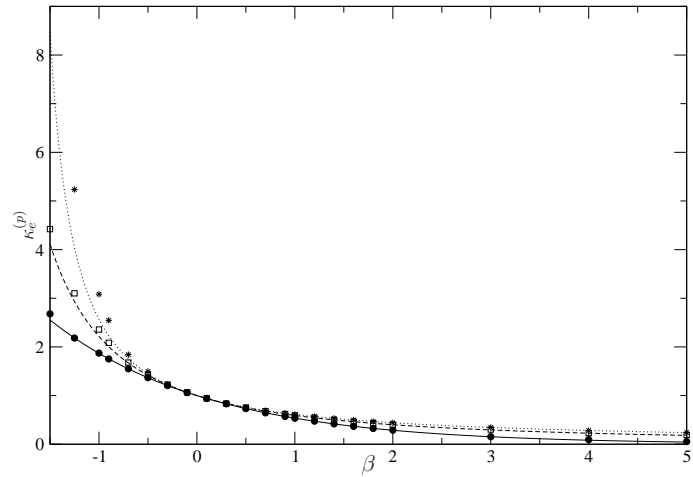}
\end{center}
\caption{Analytical calculation of equation (\ref{eqkep}) for $\kappa_e^{(p)}$ for $a=0.5$ in 2 (solid line), 
3 (dashed line) and 4 (dotted line) dimension, confronted with direct simulation of the random diffusivity problem. 
The circles stand for $2D$, squares for $3D$ and stars for $4D$. We have an excellent agreement except for $\beta<-1$ in 4D.
In this case, $\kappa_e^{(p)}$ become indeed very large which force us to use smaller time step and thus the simulations are harder.}
\label{kepa0.5}
\end{figure}

The predictions of the RG analysis can also be directly compared with a simulation of the stochastic
equation (\ref{sde}) and the results are found to be in excellent  agreement for small values of $|\beta|$.
However near the dynamical transition finite size effects play a role. Direct simulation of the gradient flow
case also  requires much longer simulation times to estimate the asymptotic diffusion constant as finite time corrections seem to be more importamt. It is clearly much better to simulate the stochastic equation ({\ref{sdek}) and then determine the  
effective diffusion constant for the gradient flow problem using the relation equation (\ref{kpkgr}).  The results using this method for the case $a=0$ are shown in figure ({\ref{ke0}). We see that the results 
for 3 and 4 dimensions are in excellent agreement with our analytical calculations and that for 
2 dimensions the only discrepancy is at large positive values of $\beta$.

\begin{figure}
\begin{center}
\epsfxsize=0.6\hsize\epsfbox{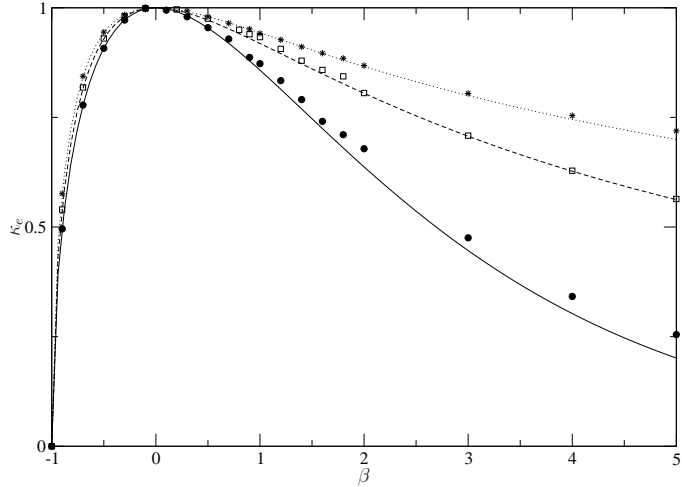}
\end{center}
\caption{Renormalization group prediction for  $\kappa_e$ for the case $a=0$ in 2 (solid line), 
3 (dashed line) and 4 (dotted line) dimensions, confronted with  numerical results deduced via the simulation of the random diffusivity  problem and using equation (\ref{kpkgr}) to estimate $\kappa_e$
($2D$ circles,  $3D$ squares and  $4D$ stars). We see that the results are excellent and improve on increasing the spatial dimension.}
\label{ke0}
\end{figure}

\section{Conclusion and Discussion}

We have seen that the renormalization group scheme developed in \cite{ddh} and \cite{dech} can be 
refined to take into account non-Gaussian potentials. The new scheme retains the merit of reproducing exactly known results in lower dimensions and has  good agreement with numerical simulations in 
cases where exact results are not known. The main breakthrough is that this scheme is capable of predicting dynamical phase transitions where the self-diffusion constant vanishes, a transition analogous to the glass transition. A key point in the analysis was to apply the approximate renormalization group not to the problem of diffusion in a random gradient field but to a mathematically related problem of diffusion in a random diffusivity field. This renormalization group calculation applied
to this problem produces a form of the celebrated Landau-Lifshitz-Matheron conjecture from the field
of random porous media as given in equation (\ref{llm}). 

Of course the problem we are considering is one with quenched disorder, in glass formers the 
disorder is thought to be somehow self induced. Let us consider for a moment the problem of
$N$ Brownian particles (of bare diffusion constant 1) interacting via a pairwise potential $u$.
The Langevin equation can be written as 
\begin{equation}
{d{\mathbf X}\over dt} = {\mathbf \eta} -\beta\nabla \nu({\bf X})
\end{equation}
Here the potential is given by
\begin{equation}
\nu({\bf X}) = \sum_{i<j} u({\mathbf X}_i-{\mathbf X}_j)
\end{equation}
i.e. the energy due to the pairwise interaction between particles. The corresponding permeability 
problem thus has
\begin{equation}
\kappa({\mathbf x}) = \exp(-\beta \nu({\mathbf x})).
\end{equation}
Clearly the system is not disordered but we shall treat it as it were and apply the formula
(\ref{llm}) to estimate the self diffusion constant. Firstly we have
\begin{equation}
\left\langle \kappa({\mathbf x})\right\rangle  = {N!\over V^N}Z_N
\end{equation}
where $V$ is the volume of the system, $N$ the number of particles (assumed indiscernable)
 and $Z_N$ is the canonical partition function for the system. Here we have simply replaced the 
disorder average by the spatial average (this is in fact the correct average to make if one
looks at the derivation of equation ({\ref{kpkgr}) \cite{rev,ddhl,ddh2} it is replaced by a disorder
average by  appealing to ergodicity). If one introduces the free energy per particle $f(\beta)$ we
find that 
\begin{equation}
\left\langle \kappa({\mathbf x})\right\rangle  = \exp\left(-N(\beta f(\beta) -\ln(\rho) +1)\right),
\end{equation}
where $\rho$ is the particle density $\rho = N/V$. Similarly we denote the dimension of the 
space of the diffusivity problem by $D= Nd$ where $d$ is the physical space dimension and find
\begin{equation}
\left\langle \kappa({\mathbf x})^{1-{2\over D}}\right\rangle^{1\over \left(1-{2\over D}\right)} = 
\exp\left[-N\left(\beta f\left(\beta\left(1-{2\over D}\right)\right) -{\ln(\rho) -1\over \left(1-{2\over D}\right)}\right)\right].
\end{equation}
We see that the two quantities above have logarithms which are extensive  in $N$ but we expect the 
self-diffusion constant to be intensive, remarkably the ratio of the two above quantities is intensive
and we find to leading order in $N$ that
\begin{equation}
\kappa_e = \exp\left({2\beta^2\over d}f'(\beta) +{2\ln(\rho)-2\over d}\right).
\end{equation}
Now we use the trivial thermodynamic identity that the entropy per particle is given by
$s=-\partial f/\partial T = \beta^2\partial f/\partial \beta$ and also the fact that at $\beta=0$ we have
$s(0) = -\ln(\rho) + 1$, to obtain
\begin{eqnarray}
\kappa_e &=& \exp\left(2{(s(\beta)-s(0))\over d}\right) \nonumber \\
&=& \exp\left({2\over d} s_{ex}(\beta)\right),
\label{part}
\end{eqnarray}
where $s_{ex}$ is simply the excess entropy per particle with respect to the perfect gas.
Thus the approximative line of mathematical reasoning we have followed has lead to a quite interesting 
relation between a dynamic quantity (the diffusion constant) and a thermodynamic quantity (the excess entropy
per particle). The celebrated Adam-Gibbs relation for glasses  relates the relaxation time to the configurational entropy \cite{glass}, here we have a relation between the self diffusion constant and
the full entropy and the relationship is quite different from the Adam-Gibbs form.   However for some time in the chemical physics literature 
\cite{ros1,ros2,dzu1,miertr} it has been observed that the diffusion constant in molecular dynamics 
simulations, when written in dimensionless form, can often be well fitted (in the denser phase) by the expression
\begin{equation}
\kappa = C \exp(As_{ex}).
\end{equation}
Now our computation is for Langevin systems so one would have to course grain a real molecular
dynamics to a Brownian level to determine $\kappa_0$ (which here we have set to 1) for the effective Langevin dynamics; we thus cannot reasonably expect to predict the value $C$ without further study.
However our naive  prediction would be $A=2/d$. Numerical simulations \cite{ros1} have revealed that 
$A$ varies quite weakly depending on the species, and it is reported that $A\approx 0.65$ for hard spheres and $A\approx  0.8$ for Lennard-Jones fluids, both in three dimensions. In another study it was 
proposed that $A=1$ \cite{dzu1} and the subject is still debated and studied (see \cite{miertr}) for a
recent review.
Here we predict  $A=2/3$ which is in intriguing agreement for the hard sphere result ! The only other analytical derivation that we are aware of for relations of the type of equation (\ref{part}) is via mode coupling theory (and thus quite different to that given here) \cite{samugo} where the effect of mixtures was also included. It will be interesting to see if this last application of our method could be refined to treat mixtures and also it should be confronted with numerical simulations of Langevin systems \cite{inpro}.

\end{document}